\documentclass[12pt]{amsart}

\usepackage{epsfig}
\usepackage{graphics}
\usepackage{amssymb}

\begin{document}


\title[]{Circumspect descent prevails in solving random constraint
satisfaction problems}

\author[]{Mikko Alava}
\address{%
Laboratory of Physics, P.O. Box 1100,
FI-02015 Helsinki University of Technology, Finland
}
\email{mikko.alava@tkk.fi}

\author[]{John Ardelius}
\address{%
SICS Swedish Institute of Computer Science AB,
SE-164 29 Kista, Sweden
}
\address{%
Department of Computational Biology, AlbaNova University Centre,
SE-106 91 Stockholm, Sweden
}

\author[]{Erik Aurell}
\address{%
Department of Computational Biology, AlbaNova University Centre,
SE-106 91 Stockholm, Sweden
}
\address{%
ACCESS Linnaeus Centre, KTH-Royal Institute of
Technology, SE-100 44 Stockholm, Sweden
}

\author[]{Petteri Kaski}
\address{%
Helsinki Institute for
Information Technology HIIT, Department of Computer Science,
University of Helsinki,
P.O.Box 68, FI-00014 University of Helsinki, Finland
}
\email{petteri.kaski@cs.helsinki.fi}

\author[]{Supriya Krishnamurthy}
\address{%
SICS Swedish Institute of
Computer Science AB, SE-164 29 Kista, Sweden
}
\address{%
School of Information and Communication Technology, KTH-Royal
Institute of Technology, SE-164 40 Kista,Sweden
}

\author[]{Pekka Orponen}
\address{
  Laboratory for Theoretical Computer Science,
  P.O. Box 5400,
  FI-02015 Helsinki University of Technology, Finland
}
\email{pekka.orponen@tkk.fi}

\author[]{Sakari Seitz}
\address{%
Laboratory for Theoretical Computer Science,
P.O. Box 5400,
FI-02015 Helsinki University of Technology, Finland
}
\email{sakari.seitz@tkk.fi}


\begin{abstract}
We study the performance of stochastic local search algorithms
for random instances of the $K$-satisfiability ($K$-SAT) problem.
We introduce a new stochastic local search algorithm, 
ChainSAT, which moves in the energy landscape of a problem
instance by {\em never going upwards} in energy.
ChainSAT is a \emph{focused} algorithm in the sense that it considers 
only variables occurring in unsatisfied clauses.
We show by extensive numerical investigations that ChainSAT and
other focused algorithms solve large $K$-SAT instances almost surely 
in linear time, up to high clause-to-variable ratios $\alpha$;
for example, for $K=4$ we observe linear-time performance well beyond
the recently postulated clustering and condensation transitions in
the solution space. The performance of ChainSAT is a surprise given
that by design the algorithm gets trapped into the first local energy 
minimum it encounters, yet no such minima are encountered.
We also study the geometry of the solution space as accessed by 
stochastic local search algorithms.
\end{abstract}

\maketitle


\section{Introduction}

\subsection{Background}

Constraint satisfaction problems (CSPs) are the industrial,
commercial and often very large-scale analogues of popular
leisure-time pursuits such as the sudoku puzzle. They can be
formulated abstractly in terms of $N$ variables $x_1,x_2,\ldots,x_N$
and $M$ constraints, where each variable $x_i$ takes a value in a
finite set and each constraint forbids certain combinations of
values to the variables. The classical example of a worst-case
intractable \cite{GaJo79} constraint satisfaction problem is the
{\em $K$-satisfiability} ($K$-SAT) problem \cite{DuGP97}, 
where each variable takes a Boolean value (either 0 or 1) and each 
constraint is a clause over
$K$ variables disallowing one out of the $2^K$ possible combinations
of values. An instance of $K$-SAT can also be interpreted
directly as a spin system of statistical physics. Each constraint
equals to a $K$-spin interaction in a Hamiltonian, and thus spins
represent the original variables; the ground states of the
Hamiltonian correspond to the solutions, that is, 
assignments of values to the variables that satisfy all
the clauses (see \cite{MMZ01}).

It was first observed in the context of $K$-SAT, and then in
the context of several other CSPs \cite{DMSZ01}, that ensembles of random
CSPs have a ``phase transition,'' a sharp change in the likelihood to
be solvable \cite{MiSL92}. Empirically, algorithms have been observed
to fail or have difficulties in the immediate neighbourhood of such
phase transition points, a fact which has given rise to a large
literature \cite{DMSZ01}.
Large unstructured CSPs are solved
either by general-purpose deterministic methods, of which the
archetypal example is the Davis-Putnam-Logemann-Loveland (DPLL)
algorithm \cite{DLL62}, or using more tailored algorithms,
such as the Survey Propagation (SP) algorithm \cite{MPZ02}
motivated by spin glass theory, or variants of stochastic 
local search techniques \cite{AaLe97,HoSt05,SKM97}.

Stochastic local search (SLS) methods are competitive on some of the
largest and least structured problems of interest \cite{KaSe07}, in
particular on {\em random} $K$-SAT instances, which are
constructed by selecting independently and uniformly at random $M$
clauses over the $N$ variables, where the parameter controlling the
satisfiability of an instance is $\alpha = M / N$, the ratio of
clauses to variables. SLS algorithms work by making successive
random changes to a trial configuration (assignment of values to the
variables) based on information about a local neighbourhood in the
set of all possible configurations. Their modern history starts
with the celebrated simulated annealing algorithm of Kirkpatrick,
Gelatt and Vecchi \cite{KGV83}. From the perspective of
$K$-SAT, the next fundamental step forward was an
algorithm of Papadimitriou \cite{Papa91}, now often called
RandomWalkSAT, which introduced the notion of {\em focusing} the random
moves to rectify broken constraints. RandomWalkSAT has been
shown, by simulation and theoretical arguments, to solve the
paradigmatic case of random 3-satisfiability up to about $\alpha =
2.7$ clauses per variable, almost surely in time linear in $N$
\cite{BaHW03,SeMo03}. A subsequent influential development occurred
with Selman, Kautz and Cohen's WalkSAT algorithm \cite{SeKC96},
which mixes focused random and greedy moves for better performance.
We have previously shown that WalkSAT and several other stochastic
local search heuristics work almost surely in linear time, up to at
least $\alpha = 4.21$ clauses per variable \cite{AA06,AGK05,SeAO05}.
In comparison, the satisfiability/unsatisfiability threshold of
random 3-satisfiability is believed to be at $\alpha = 4.267$
clauses per variable \cite{MeMZ06}.

\subsection{The present work}

The present work carries out a first systematic empirical study
of random $K$-SAT for $K=4$. Our motivation for this study is
threefold.

{\em Testing the limits of local search.}
It has been empirically observed for $K=3$ that many SLS algorithms 
have a linear-time regime, which extends to the immediate vicinity of
the phase transition point \cite{AA06,AGK05,SeAO05}.
Thus, a similar investigation for higher $K$ is warranted.
Here we focus on $K=4$.

{\em The structure of the space of solutions.}
Recent rigorous results and non-rigorous predictions from spin-glass
theory suggest that the structure of the space of solutions of a
random $K$-SAT instance undergoes various qualitative changes for
$K\geq 4$, the implications of which to the performance of
algorithms should be investigated.

M\'ezard, Mora and Zecchina \cite{MeMZ05} have shown rigorously
that for $K\geq 8$ the space of solutions of
random $K$-SAT breaks into multiple clusters separated by extensive
Hamming distance. (The Hamming distance of two Boolean vectors of 
length $N$ is the number positions in which the vectors differ divided 
by $N$.)
In more precise terms, an instance of $K$-SAT is $x$-\emph{satisfiable}
if it has a pair of solutions with normalized Hamming distance
$0\leq x\leq 1$. M\'ezard, Mora and Zecchina \cite{MeMZ05}
show that, for $K\geq 8$, there exists an interval $(a,b)$,
$0<a<b<1/2$, such that, with high probability as $N\rightarrow\infty$,
a random instance ceases to be $x$-satisfiable for all $x\in (a,b)$
at a smaller value of $\alpha$ before it ceases to be
$x$-satisfiable for some $x\in [b,1/2]$.

For $K=4$, we see no evidence of gaps in the empirical
$x$-satisfiability spectrum in the linear-time regime of SLS algorithms, 
which includes the predicted spin-glass theoretic clustering points. 
In light of the rigorous results for $K\geq 8$, this suggests that the 
cases $K=4$ and $K=8$ may be qualitatively different.
Moreover, we observe that recently predicted spin-glass-theoretic 
clustering thresholds (Krzakala et al.\ \cite{KrMR07})
have no impact on algorithm performance. This puts forth the question 
whether the energy landscape of random $K$-SAT for small $K$ is in 
some regard more elementary than has been previously believed.

{\em The structure of the energy landscape.}
In the context of random $K$-SAT it is common folklore that
SLS algorithms appear to benefit from circumspect descent in 
energy, that is, from a very conservative policy of lowering
the number of clauses not satisfied by the trial configuration.
To explore this issue further, we introduce a new SLS algorithm
which we call {\em ChainSAT}. 
It is based on three ideas: 
(1) focusing, 
(2) easing difficult-to-satisfy constraints 
    by so-called {\em chaining} moves, and 
(3) {\em never going upwards in energy}; that is, the number
of unsatisfied clauses is a non-increasing function of the
sequence of trial configurations traversed by the algorithm.

By design, ChainSAT cannot escape from a local minimum of energy
in the energy landscape. Yet, empirically ChainSAT is able to find 
a solution, almost surely in linear time, up to values 
of $\alpha$ reached by SLS algorithms that are allowed to go up 
in energy, such as the Focused Metropolis Search \cite{SeAO05}. 
This observation further supports the position that random $K$-SAT 
for small $K$ may be more elementary than has been previously believed.

\subsection{Organization of the paper}

Section 2 documents our experiments with 
the FMS algorithm on random $K$-SAT for $K=4$.
Section 3 contains an empirical investigation of
$x$-satisfiability in random $K$-SAT for $K=4$
using the FMS algorithm.
Section 4 introduces the ChainSAT algorithm
and studies its performance on random $K$-SAT for $K=4,5,6$.
Section 5 presents a few concluding remarks.

\section{Experiments with Focused Metropolis Search}

The Focused Metropolis Search (FMS) algorithm \cite{SeAO05} is given in 
pseudocode in Figure~\ref{fig:fms}. This section documents our 
experiments aimed at charting the empirical linear-time region of FMS 
on random $K$-SAT for $K=4$.

\begin{figure}
\begin{center}
\begin{tabbing}
$\qquad$\=$\quad$\=$\quad$\=$\quad$\=\kill
1:\>$S$ = random assignment of values to the variables\\
2:\>{\bf while} $S$ is not a solution {\bf do}\\
3:\>\>  $C$ = a clause not satisfied by $S$ selected uniformly at random\\
4:\>\>  $V$ = a variable in $C$ selected uniformly at random\\
5:\>\>  $\Delta E$ = change in the number of unsatisfied clauses if $V$ is flipped in $S$\\
6:\>\>  {\bf if} $\Delta E \leq 0$ {\bf then}\\
7:\>\>\>    flip $V$ in $S$\\
8:\>\>  {\bf else}\\
9:\>\>\>    {\bf with probability} $\eta^{\Delta E}$\\
10:\>\>\>\>      flip $V$ in $S$\\
11:\>\>\>    {\bf end with}\\
12:\>\>  {\bf end if}\\
13:\>{\bf end while}
\end{tabbing}
\end{center}
\caption{The Focused Metropolis Search algorithm \cite{SeAO05}.} 
\label{fig:fms}
\end{figure}

\subsection{Selecting the temperature parameter}

For $K=3$ it has already been established
that the FMS algorithm has an ``operating window''
in terms of the adjustable ``temperature'' parameter $\eta$
\cite{SeAO05}. For too large values of $\eta$, the linearity (in $N$) 
is destroyed due to too large fluctuations that keep the algorithm
from reaching low energies, and the solution. For too small values
of $\eta$, the algorithm becomes ``too greedy'' leading to a
divergence of solution times. Thus, to obtain performance linear
in $N$, it is necessary to carefully optimize the parameter $\eta$.

\begin{figure}[ht!]
\vspace*{1cm}
\begin{center}
\epsfig{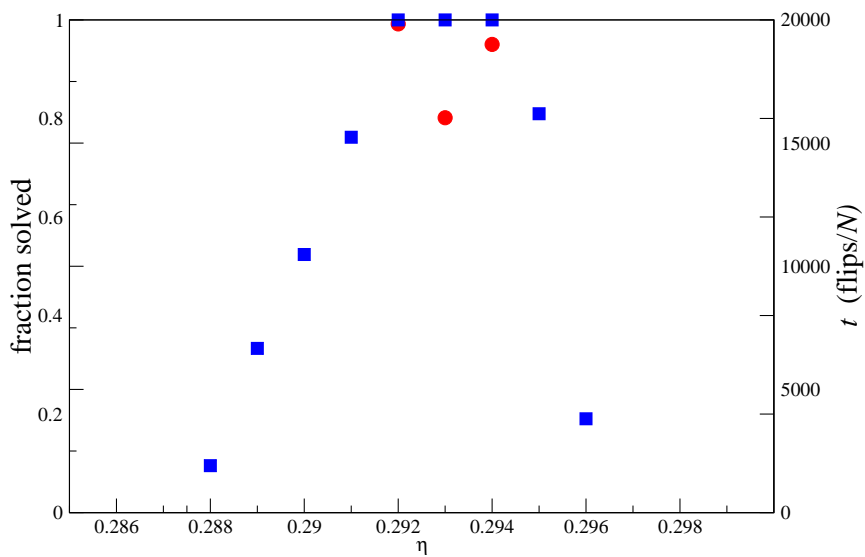}
\end{center}
\vspace*{3mm}
\caption{Optimizing the temperature parameter $\eta$ for the FMS 
algorithm on instances of random $K$-SAT at $K=4$ and $\alpha=9.6$. 
Displayed on the horizontal axis is the temperature parameter $\eta$. 
Plotted on the vertical axis is the fraction of 21 random instances 
solved within $60000\times N$ flips at $N=100000$. In the case all 21
instances are solved, also plotted is the average solution time 
(in flips/$N$). The optimum is at $\eta=0.293$. Note the 
narrowness of the operating window in terms of $\eta$. } 
\label{fig:fms-temp}
\end{figure}

Figure~\ref{fig:fms-temp} shows a typical result of the
optimization of the temperature parameter $\eta$ for random
$K$-SAT with $K=4$. Two quantities are plotted, the fraction of 
instances solved (within a threshold number of flips per variable), 
and, when all instances are solved, the corresponding average solution 
time. 

\subsection{The empirical linear-time regime of FMS}

It is evident from Figure~\ref{fig:fms-temp} that for $K=4$ 
and $\alpha=9.6$ the operating window of FMS
is already very narrow; thus it is striking that the empirical
performance of FMS is almost surely linear in $N$ within the
window. 

\begin{figure}
\vspace*{1cm}
\begin{center}
\epsfig{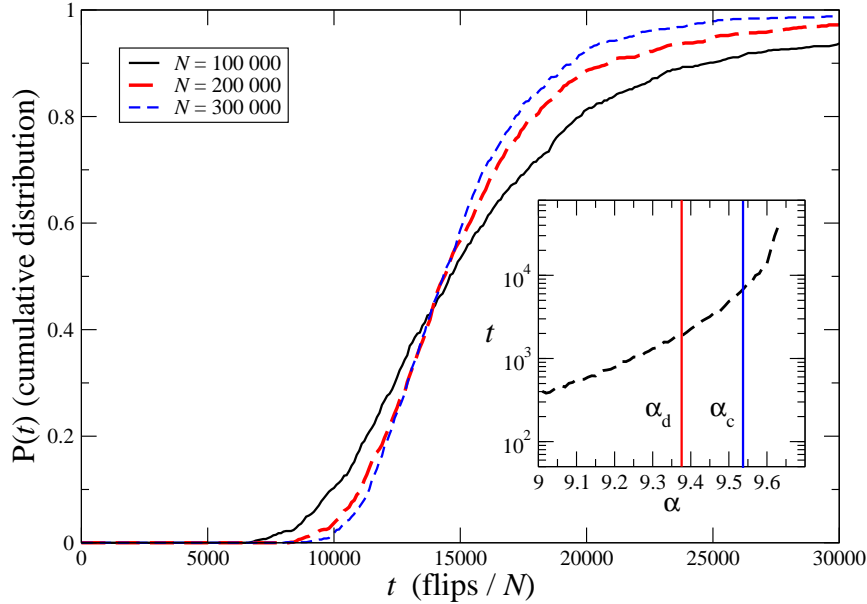}
\end{center}
\vspace*{3mm}
\caption{%
Cumulative distributions of solution times normalized by
the number of variables $N$ for the Focused Metropolis Search
algorithm \cite{SeAO05} on instances of random $K$-satisfiability 
at $K=4$ and $\alpha=9.6$.
The vertical axis indicates the fraction of 1001 random instances
solved within a given running time, measured in flips/$N$ on the
horizontal axis. 
Inset: Here we present the scaling of the algorithm 
as $\alpha$ increases (with $N=100000$).
The ``temperature''parameter of FMS is set to 
$\eta=0.293$.}
\label{fig:fms-k4-scaling}
\end{figure}

In Figure~\ref{fig:fms-k4-scaling} we present empirical evidence 
that FMS almost surely runs in time linear in $N$ for 
instances of random $K$-satisfiability with $K=4$. 
The fact that the curves get steeper
with increasing $N$ implies concentration of solution times, 
or that above-average and below-average solution times get rarer 
with $N$.
Note that the scaling implies performance almost surely linear in $N$,
and demonstrates that the linear-time regime of 
FMS extends beyond the predicted \cite{KrMR07} spin-glass theoretic
``dynamical'' and ``condensation'' transitions points.

\section{Experiments on $x$-satisfiability using FMS}

Our experimental setup to investigate $x$-satisfiability is as follows. 
For given values of $\alpha$ and $N$, we first generate a random $K$-SAT 
instance, and find one reference solution of this instance using FMS. 
Then, using FMS, we search for other solutions in the same instance.
The initial configuration $S$ for FMS is selected uniformly at random
from the set of all configurations having a given Hamming distance
to the reference solution. When FMS finds a solution, we record
the distance $x$ of the solution found to the reference solution.

\begin{figure}
\vspace*{1cm}
\begin{center}
\epsfig{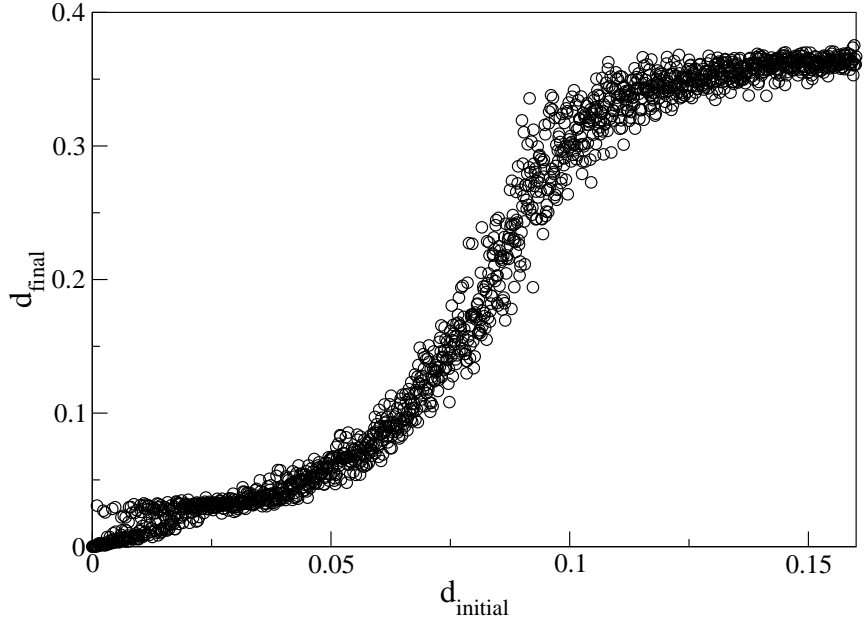}
\end{center}
\vspace*{3mm}
\caption{%
Investigation of $x$-satisfiability using FMS initialized with
a random configuration at a given Hamming distance from a
reference solution. One reference solution and one instance of
random $K$-SAT at $K=4$, $\alpha=9.6$, and $N=200000$.
The horizontal axis displays the normalized Hamming distance
of the initial configuration to the reference solution.
The vertical axis displays the normalized Hamming distance
of the solution found to the reference solution.
All of the plotted 1601 searches produced a solution, and no gaps
are visible in the vertical axis, suggesting asymptotic
$x$-satisfiability for $x\leq 0.37$.
The temperature parameter of FMS is set to $\eta = 0.293$.}
\label{fig:fms_dist1}
\end{figure}

Our experiments on random $K$-SAT for $K=4$ did not reveal any gaps 
in the $x$-satisfiability spectrum, even for $\alpha = 9.6$, 
beyond the predicted spin-glass theoretic 
``dynamical'' and ``condensation'' transitions points \cite{KrMR07}.
In particular, Figure~\ref{fig:fms_dist1} gives empirical evidence
that solutions are found at all distances smaller than the typical
distance of solutions found by FMS. This is in contrast to the
numerical results of Battaglia et al. for a balanced version of
$K=5$ \cite{Battaglia-etal05}.

Here it should be pointed out
that the solutions found by stochastic local search need {\em not} 
be typical solutions in the space of all solutions: there can be other
solutions that are not reached by FMS or other algorithms. 
Evidence of this is reflected in the ``whiteness''
status of solutions (see~\cite{SeAO05}, \cite{Pari02}, and 
Section \ref{sect:white})---all the solutions found in our experiments 
were completely white, that is, they do not have locally frozen variables.
One can of course imagine that a ``typical solution'' is not white, 
under the circumstances examined here, but as noted there is no evidence of
the existence of such.

\begin{figure}
\vspace*{1cm}
\begin{center}
\epsfig{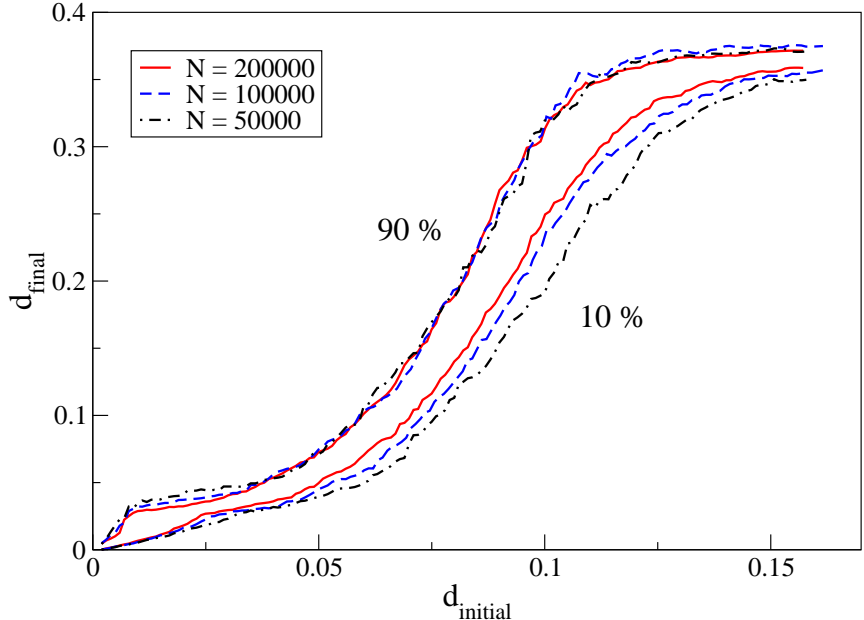}
\end{center}
\vspace*{3mm}
\caption{%
Scaling of $x$-satisfiability data obtained using FMS
on random $K$-SAT with increasing $N$.
The parameters $K=4$, $\alpha=9.6$, and $\eta=0.293$ are fixed.
The plotted 10- and 90-percentile curves are calculated from five
random instances and reference solutions for each $N=50000,100000,200000$,
with a moving window size of 0.004 in the horizontal axis.
The distances appear to converge close to the 90\%-curves.}
\label{fig:fms_dist2}
\end{figure}

Figure~\ref{fig:fms_dist2} summarizes the results of a
scaling analysis with increasing $N$ over five random instances and
reference solutions. The distance distributions appear to converge
to some specific curve without vertical sections, the absence of which
suggests that the $x$-satisfiability spectrum has no gaps below the
typical distance of solutions found by FMS in the limit of 
infinite $N$.

\begin{figure}
\vspace*{1cm}
\begin{center}
\epsfig{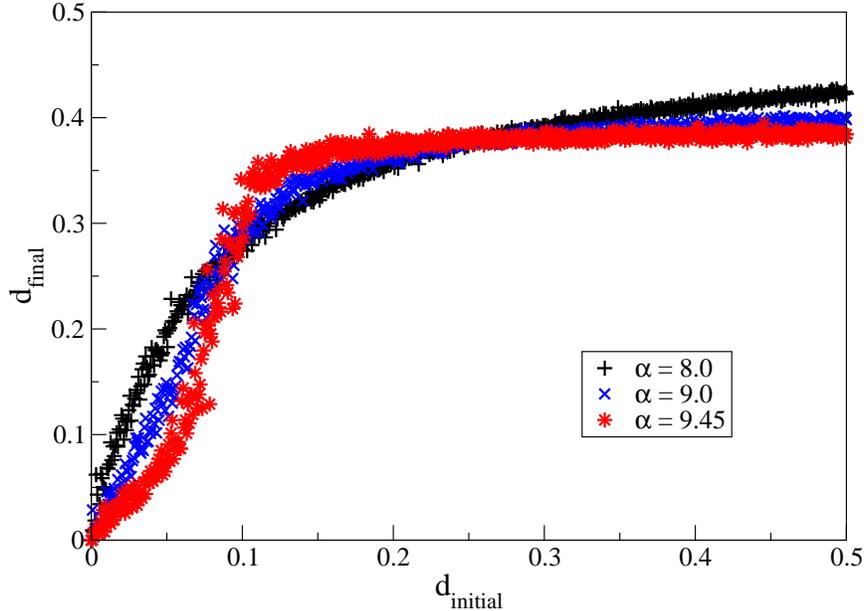}
\end{center}
\vspace*{3mm}
\caption{%
Scaling of $x$-satisfiability data obtained using FMS
on random $K$-SAT with increasing $\alpha$.
The parameters $K=4$, $N=100000$, and $\eta=0.293$ are fixed.
One random instance and one reference solution for each
$\alpha=8.0,9.0,9.45$; see Figure \ref{fig:fms_dist1}
for $\alpha=9.6$. The value $\alpha = 9.45$ is
between the predicted locations of the dynamical and the
condensation transition points \cite{KrMR07}.
No clear gap in distances is discernible in any of the cases.}
\label{fig:fms_dist3}
\end{figure}

Figure~\ref{fig:fms_dist3} summarizes the results of a
scaling analysis with increasing $\alpha$. We see that the typical
distance between solutions found by FMS decreases with increasing
$\alpha$, and that no clear gaps are apparent in the distance data.

\section{Experiments with ChainSAT}

A new heuristic which never moves up in energy is here shown to
solve random $K$-satisfiability problems almost surely in time linear
in $N$, for $K=4,5,6$.

\subsection{The ChainSAT algorithm}

Our new heuristic, ChainSAT, is given in pseudocode 
in Figure~\ref{fig:chainsat}. The algorithm (a) never increases the 
energy of the current configuration $S$; and (b) exercises circumspection in
decreasing the energy. In particular, moves that decrease the energy
are taken only sporadically compared with equi-energetic moves and
chaining moves. The latter are designed to alleviate critically
satisfied constraints by proceeding in ``chains'' of
variable-clause-variable until a variable is found which can be
flipped without increase in energy.  Focusing is employed for the
non-chaining moves. The structure of ChainSAT has the basic idea of
helping to flip a variable to satisfy an original broken constraint.

\begin{figure}
\begin{center}
\begin{tabbing}
$\qquad$\=$\quad$\=$\quad$\=$\quad$\=\kill
1:\>$S$ = random assignment of values to the variables\\
2:\>chaining = FALSE\\
3:\>{\bf while} $S$ is not a solution {\bf do}\\
4:\>\>  {\bf if not} chaining {\bf then}\\
5:\>\>\>    $C$ = a clause not satisfied by $S$ selected uniformly at random\\
6:\>\>\>    $V$ = a variable in $C$ selected u.a.r.\\
8:\>\>  {\bf end if}\\
9:\>\> $\Delta E$ = change in the number of unsatisfied clauses if $V$ is flipped in $S$\\
10:\>\>  chaining = FALSE\\
11:\>\>  {\bf if} $\Delta E=0$ {\bf then}\\
12:\>\>\>    flip $V$ in $S$\\
13:\>\>  {\bf else if $\Delta E < 0$}\\
14:\>\>\>    {\bf with probability} $p_1$\\
15:\>\>\>\>       flip $V$ in $S$ \\
16:\>\>\>    {\bf end with}\\
17:\>\>  {\bf else}\\
18:\>\>\>    {\bf with probability} $1-p_2$\\
19:\>\>\>\>       $C$ = a clause satisfied only by $V$ selected
                        u.a.r.\\
20:\>\>\>\>       $V'$ = a variable in $C$ other than $V$ selected 
                         u.a.r.\\
21:\>\>\>\>       $V$ = $V'$\\
22:\>\>\>\>       chaining = TRUE\\
23:\>\>\>    {\bf end with}\\
24:\>\>  {\bf end if}\\
25:\>{\bf end while}
\end{tabbing}
\end{center}
\caption{The ChainSAT algorithm.} \label{fig:chainsat}
\end{figure}

The ChainSAT algorithm has two adjustable parameters, one ($p_1$)
for controlling the rate of descent (by accepting energy-lowering
flips) and another ($p_2$) for limiting the length of the chains to
avoid looping. We omit data related to the optimization of these 
parameters since the procedure is simply an empirical (vary the 
parameters, check outcome), similar to one documented for the FMS 
algorithm in Figure~\ref{fig:fms-temp}.

\subsection{ChainSAT performance}

\begin{figure}
\vspace*{1cm}
\begin{center}
\epsfig{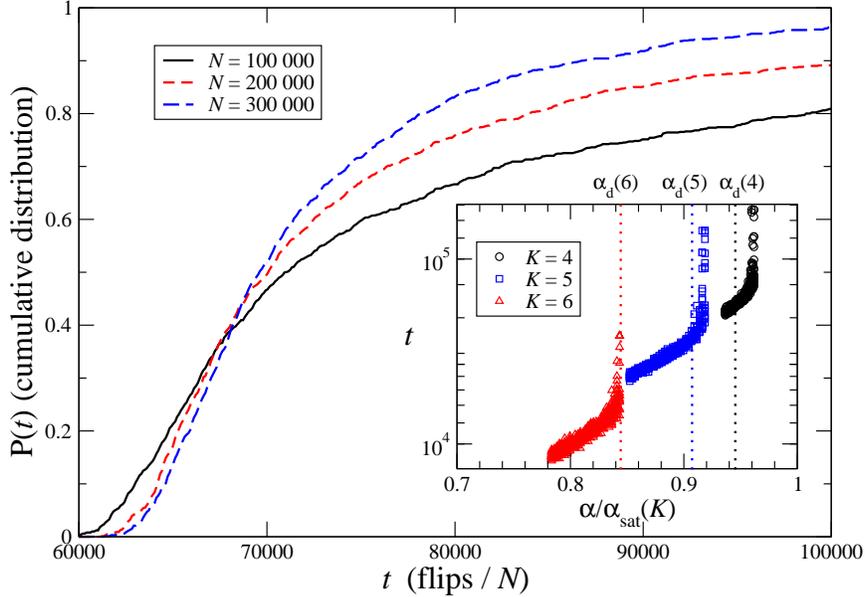}
\end{center}
\vspace*{3mm}
\caption{Cumulative distributions of solution times normalized by number
of variables $N$ for the ChainSAT algorithm on random $K$-satisfiability
instances at $K=4$ and $\alpha = 9.55$. The vertical axis indicates 
the fraction of 1001 random input instances solved within a given 
running time, measured in flips/$N$ on the horizontal axis. 
Inset: Here we present the scaling of the algorithm for $K=4,5,6$ at
$N=100000$ with increasing $\alpha$;
the values of $\alpha(K)$ in the horizontal axis have
been normalized with $\alpha_{\mathrm{sat}}(K)$,
which has the empirical values
$\alpha_{\mathrm{sat}}(4)=9.931$,
$\alpha_{\mathrm{sat}}(5)=21.117$, and
$\alpha_{\mathrm{sat}}(6)=43.37$ \cite{MeMZ06}.
The parameters of ChainSAT have been chosen to be small enough 
to work at least up to the predicted ``dynamical transition'' \cite{KrMR07}: 
we have set 
$p_1=p_2=0.0001$ ($K=4$), $0.0002$ ($K=5$), and $0.0005$ ($K=6$).}
\label{fig:chainsat-scaling}
\end{figure}

In Figure~\ref{fig:chainsat-scaling} we present empirical evidence 
that ChainSAT almost surely runs in time linear in $N$ for random 
$K$-satisfiability problems with $K=4,5,6$. 
The fact that the curves get steeper
with increasing $N$ implies concentration of solution times, 
or that above-average and below-average solution times get rarer 
with $N$.

Since the algorithm
never goes uphill in the energy landscape, local energy minima 
cannot be an obstruction to finding solutions, at least in the region 
of the energy landscape visited by this algorithm. On the other hand, 
when ChainSAT fails to find a solution in linear time, this can also
result from simply getting lost---in particular, the fraction of 
moves that lower the energy over those that keep it 
constant may dwindle to zero.

\subsection{Whiteness}
\label{sect:white}

\begin{figure}[ht!]
\vspace*{1cm}
\begin{center}
\epsfig{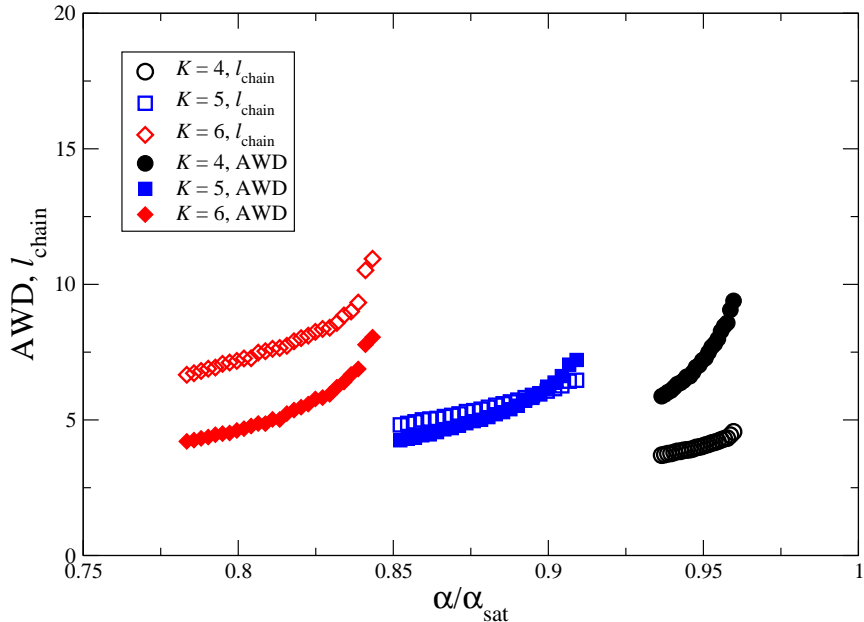}
\end{center}
\vspace*{3mm}
\caption{%
The average chain length in ChainSAT and the average whiteness depth
of the solutions found in random $K$-SAT for $K=4,5,6$.
Each plotted value is the average over 21 random instances.
The values of $\alpha(K)$ in the horizontal axis have
been normalized with $\alpha_{\mathrm{sat}}(K)$,
which has the empirical values
$\alpha_{\mathrm{sat}}(4)=9.931$,
$\alpha_{\mathrm{sat}}(5)=21.117$, and
$\alpha_{\mathrm{sat}}(6)=43.37$ \cite{MeMZ06}.
The ChainSAT parameters are set to
$p_1=p_2=0.0001$ ($K=4$), $0.0002$ ($K=5$), and $0.0005$ ($K=6$).
}
\label{fig:chain}
\end{figure}

To provide a further empirical analysis of ChainSAT, we next present
Figure~\ref{fig:chain}. This is discussed not in terms of solution
times and the range of $\alpha$ achieveable with a bit of tuning,
but in terms of two quantities: (i) the average chain
length $l_{\mathrm{chain}}$ during the course of finding a solution
and (ii) the average whiteness depth (AWD). In more precise terms,
the average chain length is $l_{\mathrm{chain}}=f/m-1$, where $f$ is
the total number of iterations of the main loop of ChainSAT and $m$
is the number of times the if-statement controlled by the chaining
flag in the main loop is executed.

The AWD is related to the result of the so-called whitening
procedure \cite{Pari02}, described in pseudocode in Figure
\ref{fig:whitening}, that is applied to the solution found when
ChainSAT terminates. The whiteness depth of a variable is defined as 
the value of $D$ in the whitening procedure at the time the variable gets
marked (whitened); the value is infinite if the variable never gets
marked (whitened) during the whitening procedure. The AWD of a
solution is the average of the whiteness depths of the variables.
See \cite{SeAO05} for an empirical discussion of AWD in the context
of random $K$-SAT for $K=3$. The key observation here is that the 
solutions found by ChainSAT all have a finite AWD. This in loose 
terms means that there is ``slack'' in the solution.

\begin{figure}
\begin{center}
\begin{tabbing}
$\qquad$\=$\qquad$\=$\qquad$\=$\qquad$\=\kill
1:\>initially all clauses and variables are unmarked (non-white)\\
2:\>mark (whiten) every clause that is unsatisfied\\
3:\>mark (whiten) every clause that has more than one true literal\\
4:\>$D$ = $0$\\
5:\>{\bf repeat}\\
6:\>\>mark (whiten) any unmarked variables that appear as satisfying\\
  \>\>\ \ literals only in marked clauses\\
7:\>\>{\bf if} all the variables are marked {\bf then}\\
8:\>\>\>declare that $S$ is completely white\\
9:\>\>\>{\bf halt}\\
10:\>\>{\bf end if}\\
11:\>\>{\bf if} no new variables were marked in this iteration {\bf then}\\
12:\>\>\>declare that $S$ has a core\\
13:\>\>\>{\bf halt}\\
14:\>\>{\bf end if}\\
15:\>\>mark (whiten) any unmarked clauses that contain at least\\
   \>\>\ \ one marked variable\\
16:\>\>$D$ = $D+1$\\
17:\>{\bf end repeat}\\
\end{tabbing}
\end{center}
\caption{The whitening algorithm for a configuration $S$.}
\label{fig:whitening}
\end{figure}

Based on Figure~\ref{fig:chain} it is clear that 
increasing the value of $\alpha$ has the same 
effect for $K=4,5,6$: 
the average chain length $l_{\mathrm{chain}}$ increases, 
and so does the AWD. Note that the ratio AWD/$l_{\mathrm{chain}}$ 
increases with $\alpha$.

\section{Concluding remarks}

We have here shown empirically that local search heuristics can be
designed to avoid traps and ``freezing'' in random 
$K$-satisfiability, with solution times scaling linearly in $N$. 
This requires that circumspection is exercised---too greedy a descent 
causes the studied algorithms to fail for reasons unclear. 
A physics inspired interpretation is that during a run the algorithm 
has to ``equilibrate'' on a constant energy surface.

In terms of the parameter $\alpha$, it is the pertinent question as to
how far the ``easy'' region from which one finds these 
solutions extends. For small $K$ it may be possible that this is true 
all the way to the satisfiability/unsatisfiability transition point. 
The empirical evidence we have here presented points 
towards a divergence of the prefactor of the linear scaling 
in problem size well below $\alpha_{sat}$. Furthermore, this
divergence is stronger for higher values of $K$. For large values of $K$,
the absence of traps may however in any case be considered unlikely,
as the rigorous techniques used to show clustering of
solutions for $K\geq 8$ \cite{MeMZ05} can also be used to show that 
there exist pairs of distant solutions separated by an extensive energy
barrier from each other. This suggests also the existence of local
minima separated by extensive barriers. On the other hand, our 
present results for small $K$ give no evidence in this direction. 
In particular, for $K=4$ we have shown empirically that the energy 
landscapes can be navigated with simple randomized heuristics beyond 
all so far predicted transition points, apart from the
satisfiability/unsatisfiability transition itself.

Our experiments also strongly suggest that the space of solutions for 
$K = 4$ at least up to $\alpha = 9.6$ does not break into multiple 
clusters separated by extensive distance. All the solutions found have 
``slack'' in the sense that they have a finite AWD.
Is there an efficient way to find solutions that are not ``white'' 
in this sense; put otherwise, is the existence of ``white'' solutions
necessary for ``easy'' solvability? 

All these observations present further questions about the
structure of the energy landscape, the solution space, and
the workings of algorithms for random CSPs.
They also leave us with challenges and constraints to
theoretical attempts to understand these, including approaches
from the physics of spin glasses.

{\bf Acknowledgements:} This work was
supported by the Integrated Project EVERGROW of the European Union
(J.A and S.K), by the Swedish Science Council through Linnaeus
Centre ACCESS (E.A.), and by the Academy of Finland under Grant
117499 (P.K.) and through the Center of Excellence Program (M.A. and
S.S.). We thank the Department of Computer Science of University of
Helsinki and SICS for the use of a little over 6 years of CPU time.


\end{document}